\documentstyle[twocolumn,prb,aps,epsfig]{revtex}

\begin{document}
\title{Dzyaloshinskii-Moriya interaction in NaV$_2$O$_5$: \\
       a microscopic study.}

\author{Roser Valent\'\i$^1$, Claudius Gros$^1$ and Wolfram Brenig$^2$} 

\address{$^1$Fachbereich Physik, University of the Saarland,
66041 Saarbr\"ucken, Germany.}

\address{$^2$Institut f\"ur Theoretische Physik, 
TU-Braunschweig, Mendelsohnstr. 3, 38106 Braunschweig, Germany.}

\date{\today}
\maketitle

\begin{abstract}
We present a unified account of magnetic exchange and Raman scattering in
the quasi-one-dimensional transition-metal oxide NaV$_2$O$_5$. Based on a
cluster-model approach explicit expressions for the exchange integral and
the Raman-operator are given.  It is demonstrated that a combination of
the electronic-structure and the Dzyaloshinskii-Moriya interaction,
allowed by symmetry in this material, are responsible for the finite Raman
cross-section giving rise to both, one- and two-magnon scattering
amplitudes.
\end{abstract}


\section{INTRODUCTION}
More than four decades ago, Dzyaloshinskii \cite{Dzyalo58} and Moriya
\cite{Moriya60} showed that the inclusion of spin-orbit coupling into the
description of low-symmetry magnetic systems  generates an anisotropic
exchange interaction, the so-called Dzyaloshinskii-Moriya (DM)
interaction.

In the early nineties, this interaction was discussed intensively in
connection with the copper-oxide compounds. In particular, La$_2$CuO$_4$
exhibits a small gap in the spin-wave spectrum and a finite net
ferromagnetic moment in each plane due to an out-of-plane canting of the
spins. These features were attributed to DM interactions.
\cite{Coffey90,Coffey91,Shekhtman92} Yildirim {\it et al.}
\cite{Yildirim95} did a careful microscopic study of this me\-chanism
for tetragonal copper-oxide systems. In particular, their analysis proved
that the orthorhombic distortion present in these materials is irrelevant to
the out-of-plane magnetic anisotropy.  Moreover, they showed that not only
the antisymmetric anisotropic superexchange between two neighboring spins
is important but the symmetric one as well.
\cite{Shekhtman92,Yildirim95,Shekhtman93,Kaplan83}
  
The DM interaction has gained renewed interest in the context of the novel
transition-metal oxide NaV$_2$O$_5$, which is believed to be a
quarter-filled ladder compound in its high-temperature phase.
\cite{Smolinski98} At $T_C=34\,\mbox{K}$ a phase transition, the interpretation
of which is still controversial,  takes place in this material where
charge ordering ($2V^{+4.5}\rightarrow V^{+4}+V^{+5}$) occurs
simultaneously with the opening of a spin-gap of approximately
$10\,\mbox{meV}$.\cite{Isobe96} A series of recent studies has addressed
the nature of the low-temperature state
\cite{Gros99,Luedecke99,Boer99,Smaalen99,Gros00,Ohama00}.

In this context it is of interest that very recent Electron Spin Resonance
(ESR) experiments \cite{Luther98,Lohmann00} have detected a considerable
anisotropy of the absorption intensity with respect to the magnetic field
orientation, which has been attributed to the DM interaction. Apart from
ESR,  Raman scattering in the presence of a magnetic field is an alternative
experiment for the observation of possible effects due to DM interactions.
Unfortunately however, at present, the various experimental settings in
search for such effects in Raman scattering have been unsuccessful for
NaV$_2$O$_5$. \cite{Lemmens}  This may not be conclusive yet, since the
Raman cross section of the relevant scattering process could be too small
to be observable.

In order to shed some light onto this scene, a microscopic analysis of the
magnetic exchange and Raman scattering-operator seems highly desirable.
However, apart from early work specific to the copper-oxides
superconductors\cite{Brenig} such analysis is lacking. Therefore, it is
the purpose of this paper to present a detailed description of the Raman
operator for NaV$_2$O$_5$. In this context particular emphasis will be
given to the role of  the DM interaction which by symmetry is allowed in
this material.

\begin{figure}[t]
\centerline{
\epsfig{file=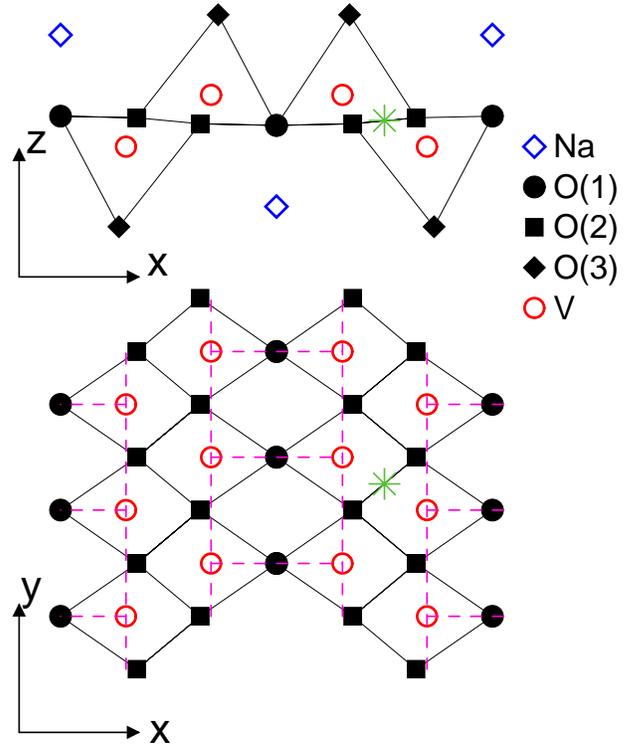,width=0.45\textwidth} 
}
\vspace{4pt}
\caption{\label{Fig1}
Crystal structure of NaV$_2$O$_5$ in the high-tempera\-ture
phase. The star denotes
the location of the center of inversion, the dashed
lines the constituting V-O-V ladders. }
\end{figure}


\section{HAMILTONIAN}

Above the spin-charge transition temperature T$_{C} $ NaV$_2$O$_5$
crystallizes in the centrosymmetric Pmmn space
group.\cite{Smolinski98,Schnering98}  The compound consists of VO$_5$
square pyramids sharing edges in the $ab$ layer and chains of Na located
between the $ab$ layers. The superexchange interaction between vanadium
sites is mediated through the pyramid's base oxygens and the relevant
structural element of NaV$_2$O$_5$ can be thought of consisting of ladders
of V-O-V rungs along $b$ which are weakly coupled along $a$  (see Fig.\
\ref{Fig1}).

Discarding single-ion anisotropy, a general form of any scalar
two-spin interaction between consecutive rungs along the $b$-direction of
the ladder in NaV$_2$O$_5$  consists of two contributions
\begin{equation}
H^{(S)}=J\,H^{(Heis)}+D\,H^{(DM)}
\end{equation}
i.e. the isotropic Heisenberg exchange $H^{(Heis)}$ and the
Dzyaloshinskii-Moriya interaction $H^{(DM)}$:
\begin{eqnarray}
H^{(Heis)}=\sum_l {\bf S}_l\cdot{\bf S}_{l+1}, \nonumber \\
H^{(DM)}=\sum_l {\bf e}\cdot 
\left({\bf S}_l\times{\bf S}_{l+1}\right)
\label{hamiltot}
\end{eqnarray}
where ${\bf S}_l$ denotes the total spin on rung $l$. The form of the
Dzyaloshinskii-Moriya (DM) vector ${\bf e}$ is determined by requiring
that the energy of any configuration of spins has to be invariant under
the symmetry transformations of the crystal structure. In our case,
crystallography allows for a DM vector along $c$, i.e. ${\bf e}=(0,0,1)$.
Note that this vector is defined {\it locally}  in each unit cell  and is
not forbidden by the inversion center of the crystal structure, which lies
in between two V-O-V ladders (see Fig.\ \ref{Fig1}).

The standard derivation of (\ref{hamiltot}) for the case of magnetic
moments {\em localized} at single ionic sites can be found in the
literature \cite{Moriya60,Shekhtman93,Kaplan83} where it is shown that
both terms in this Hamiltonian can be derived from generalized exchange
processes. In the present paper, however, we aim at a microscopic
derivation of (\ref{hamiltot}) for the case of the {\em mixed-valence}
system NaV$_2$O$_5$. We start from a three-band Hubbard-model for
NaV$_2$O$_5$ in which we retain only the two active $d$-orbitals of the V-
and the single O(1)-site on each rung. For simplicity we consider the
O(2)-sites on the legs of the ladder to be integrated out, giving rise to
an effective V-V hopping $-t_\parallel$ along the leg. We denote by $d_{l
\alpha \sigma}^\dagger$ the creation operators for spin-$\sigma$ electrons
in V-$d_{xy}$ orbitals on site $\alpha$ ($\alpha=1,2$) of the $l$'th rung
and by $p_{l \sigma}^\dagger$ the creation operator of spin-$\sigma$
electrons in the O(1)-$p_y$ orbital on the $l$'th rung.  The Hamiltonian
$H=H_0+H_1$ reads:
\begin{eqnarray}\label{r3}
H_0&=&\sum_{l, \sigma}
\left( t_{\perp} d_{l 1 \sigma}^\dagger d_{l 2 \sigma}^{\phantom{\dagger}}
   + t_{dp}(d_{l 1 \sigma}^\dagger + d_{l 2 \sigma}^{\dagger})p_{l \sigma}
 + c.c.
\right)   \nonumber \\
&+&\epsilon_p\sum_{l\sigma}
p_{l\sigma}^\dagger p_{l\sigma}^{\phantom{\dagger}} 
 + U\sum_{l\alpha}
d_{l\alpha\uparrow}^\dagger d_{l\alpha\uparrow}^{\phantom{\dagger}}
d_{l\alpha\downarrow}^\dagger d_{l\alpha\downarrow}^{\phantom{\dagger}}
\nonumber \\
H_1&=&\sum_{l,\alpha\sigma}\left(t_\sigma ~
d_{l\alpha\sigma}^\dagger d_{l+1\alpha\sigma}^{\phantom{\dagger}}
+ t_\sigma^* ~
d_{l+1\alpha\sigma}^\dagger d_{l\alpha\sigma}^{\phantom{\dagger}}
                            \right) 
\nonumber\\
&-&t_{pp}\sum_{l,\sigma}\left(
p_{l\sigma}^\dagger p_{l+1\sigma}^{\phantom{\dagger}}
+p_{l+1\sigma}^\dagger p_{l\sigma}^{\phantom{\dagger}}
                                 \right)~,
\label{H_1}
\end{eqnarray}
The hybridization matrix elements in Eq.\ (\ref{H_1}) are: (i) $t_{\perp}$
which denotes the direct hopping of electrons between the V-$d_{xy}$
orbitals on sites 1 and 2 on a rung, (ii) $t_{dp}$ which denotes the
charge transfer integral between the V-$d_{xy}$ and O(1)-$p_y$ orbitals
on a rung,  (iii) $t_\sigma= -t_\parallel+i\sigma \tilde{\lambda}$ is the
hopping of electrons with spin $\sigma=\pm1$ between the V-$d_{xy}$
orbitals on two consecutive rungs along the ladder direction $b$. The
spin-dependence of this hopping-integral is allowed due to the lack of a 
center of inversion in-between two rungs and arises from the spin-orbit
coupling\cite{Moriya60} of strength $\lambda$, where $\tilde{\lambda}
\sim\lambda$. The transfer matrix-elements $t_\sigma$ are
diagonal in the spin quantum numbers because we have chosen the
quantization axis for the spin to be along $c$, i.e. the main
crystallographic axis, and finally  (iv) $t_{pp}$ denotes the hopping of
electrons between O(1)-$p_y$ orbitals on two consecutive rungs along $b$.
The spin-dependence of $t_{pp}$ is small and will be discarded in the
remainder of this paper.

The parameters involved in (\ref{H_1}) have been estimated
\cite{Smolinski98} to be $\epsilon_p\approx-3\,\mbox{eV}$, $t_{\perp}
\approx 0.25\, \mbox{eV}$, $t_{dp}\approx-1\,\mbox{eV}$,
$t_\parallel\approx-0.175\,\mbox{eV}$,
$t_{pp}\approx0.5\,\mbox{eV}$ and $U\approx2.8\,\mbox{eV}$. The
Coulomb repulsion $U$ leads to the formation of local moments on the rungs
and in the following we will study the interaction between these local
moments. For simplicity we consider the case $U\rightarrow\infty$, since
we expect that any finite $U$ will lead to qualitatively similar results
while increasing the complexity of the calculation needlessly.\cite{Sa00}


\section{EXCHANGE COUPLINGS}
As a first step towards the evaluation of the exchange couplings $J$ and
$D$, we diagonalize the hamiltonian $H_0$ for an isolated rung. The
ground state belongs to the 3-particle subspace. In addition, for the
calculation of the exchange matrix elements, intermediate states in the 2-
and 4-particle sector on a rung are required.  Details of the derivation
of the relevant eigenstates and eigenvalues of this cluster problem are
stated explicitly in Appendix A. The exchange matrix elements are obtained
by considering the process which describes a spin-flip between two
consecutive rungs up to second order perturbation theory in $H_1$,  as
defined by $J^{-+}$ and $J^{+-}$ in Eq.\ (\ref{ap10}). In particular,
\begin{eqnarray}
J=\frac{1}{2}(J^{-+}+J^{+-})=Re(J^{-+}).
\label{J}
\end{eqnarray}
Substituting the
electronic model-parameters cited in the previous section into Eqns.\
(\ref{ap11}) and (\ref{ap14}) we obtain $J \approx 0.049 eV  \approx 568 K
$ which agrees very well with the experimental value \cite{Isobe96,Mila96}
of $J_b \approx 560K$.
 
Noting that $t_\downarrow=t_\uparrow^*$ and that ${\bf D}\cdot \left({\bf
S}_l\times{\bf S}_{l+1}\right)=(D/2i)\left(S_l^- S_{l+1}^+ - S_l^+
S_{l+1}^-\right)$, we have
\begin{eqnarray}
D= \frac{1}{2i}(J^{-+}-J^{+-})=Im(J^{-+})
\label{D}
\end{eqnarray}
for the DM-coupling. Moriya \cite{Moriya60} has estimated that the order of
magnitude of $D$ should be
\begin{eqnarray}
D \sim (\Delta g/g) J
\label{moriyaD}
\end{eqnarray}
where $g$ is the gyromagnetic ratio of the Vanadium ion in octahedral
crystal symmetry and $\Delta g$ is the corresponding deviation from the
free-electron value. By considering the $g$-values obtained from ESR
measurements\cite{Lohmann00}, we arrive at $ \Delta g/g \approx 0.01$.
Then, from Eq.\ (\ref{D}), Eq.\ (\ref{moriyaD}) and Eq.\ (\ref{ap14})
where we evaluate $D$ as a function of $\tilde{\lambda}$, we get an
estimate for $\tilde{\lambda}$, i.e. $\tilde{\lambda}\approx 1 meV$.


\section{RAMAN SCATTERING}
Fleury and Loudon \cite{Fleury68} have shown that light scattering from a
spin system, depending on the polarization geometry of the incoming and
outgoing electric fields, can lead to inelastic photon-induced
superexchange.  This has established Raman scattering as an important
probe to obtain information on the local exchange dynamics in magnetic
systems complementary to inelastic neutron scattering (INS). In the following
we will generalize the early ideas of Fleury and Loudon to the case of
NaV$_2$O$_5$ clarifying the role of the DM interaction. In particular we
find that in the case of a polarization of both, the incoming and outgoing
photon fields parallel to $b$, i.e. along the legs of the ladder, the Raman
scattering operator $H^{(R)}$ can be expressed as
\begin{eqnarray}
H^{(R)}(\omega_{in,out}) = && J_R(\omega_{in,out}) H^{(Heis)}
\nonumber\\
&& + D_R(\omega_{in,out}) H^{(DM)}
\label{H_R}
\end{eqnarray}
where the $\omega_{in}$  and $\omega_{out}$ are the frequencies of the
incoming and outgoing photons.  The microscopic derivation
$H^{(R)}(\omega_{in,out})$ is placed into the appendix. It is identical to
that of the magnetic exchange integral with however the virtual hopping
into the intermediate state of the exchange process driven by the coupling
of the vector potential ${\bf A}=(0,A_b,0)$ to the current
operator\cite{Shastry9091}, i.e. $H_1$ of (\ref{r3}) has to be replaced by
${\bf j}\cdot{\bf A}$ with the current operator ${\bf j}=(j_a,j_b,j_c)$,
\begin{eqnarray*}
j_b=i \mbox{e}\sum_{l,\sigma}\Big(
t_{pp}\left[
 p_{l\sigma}^\dagger p_{l+1\sigma}^{\phantom{\dagger}}
-p_{l+1\sigma}^\dagger p_{l\sigma}^{\phantom{\dagger}}\right]
\qquad\\ 
+\sum_\alpha\left[ t_\sigma
 d_{l\alpha,\sigma}^\dagger d_{l+1\alpha,\sigma}^{\phantom{\dagger}}
-t_\sigma^*
 d_{l+1\alpha,\sigma}^\dagger d_{l\alpha,\sigma}^{\phantom{\dagger}}
           \right]   \Big)~.
\end{eqnarray*}
The total magnetic Raman  scattering amplitude is then given up to second
order in $j_b/e$ by Eq.\ (\ref{ap101}) of Appendix A. From this,
the definition of $J_R(\omega_{in,out})$ and $D_R(\omega_{in,out})$
is analogous to (\ref{J}) and
(\ref{D})
\begin{eqnarray}
J_R(\omega_{in,out})= && Re(R^{-+}(\omega_{in,out}))
\nonumber\\ 
D_R(\omega_{in,out})= && Im(R^{-+}(\omega_{in,out}))
\end{eqnarray}

Note that a magnetic Raman process is possible only if $H^{(R)}(
\omega_{in,out})$ induces transitions between different eigenstates
of $H^{(S)}$ \cite{Muthu96,Singh96,Brenig97}, i.e. if
\begin{eqnarray}\label{commu}
[H^{(R)}(\omega_{in,out} && ), H^{(S)}] = ( J_R(\omega_{in,out}) D  
\nonumber \\
&& -D_R(\omega_{in,out})J)\,[H^{(Heis)},H^{(DM)}] \ne0
\label{conmutator}
\end{eqnarray} 
>From (\ref{commu}) we
conclude that magnetic Raman scattering from NaV$_2$O$_5$, if modeled by
(\ref{r3}), arises because two conditions are simultaneously satisfied.
First, the existence of a spin-orbit coupling leads to a non-vanishing
commutator in (\ref{commu}).  Second, because the number of available
paths for the magnetic and the photon induced exchange is larger than one
{\em and} because $\omega_{in,out}\neq 0$ the factor of
$J_R(\omega_{in,out}) D-D_R(\omega_{in,out})J$ is nonzero.  The latter is
true despite the formal similarity between the Raman scattering amplitude
and the magnetic exchange integral, because  $H^{(R)}(\omega_{in,out})$
displays an additional dependence on the photon energies. More
specifically, for a {\em single} exchange path $J_R(\omega_{in,out})
D-D_R(\omega_ {in,out})J=0$ for any value of $\omega_{in,out}$ while for
more than one exchange path $J_R(\omega_{in,out}) D-D_R(\omega_{in,out})J$
vanishes only at $\omega_{in,out}=0$.

Next, we would like to point out that from the two terms, $H^{(Heis)}$ and
$H^{(DM)}$ which make up $H^{(R)}(\omega_{in,out})$ it is actually
$H^{(DM)}$ which drives the magnetic Raman process. Up to now we have only
considered anisotropic contributions to $H^{(S)}$ to leading order in
$\tilde{\lambda}$.  Kaplan \cite{Kaplan83} and Shekhtman {\it et al}.
\cite{Shekhtman93} have shown that, in general, the next-order term
$H^{(KSAE)}\sim\sum_l ({\bf e}\cdot{\bf S}_l) ({\bf e}\cdot{\bf S}_{l+1})$
contributes to the spin Hamiltonian with a very specific prefactor:
\begin{eqnarray}
H^{(S)}=J\,H^{(Heis)}+D\,H^{(DM)}  \nonumber \\
+\left(\sqrt{J^2-D^2}-J\right)\,H^{(KSAE)}.
\end{eqnarray}
Using this it is possible to transform $H^{(S)}$ into an equivalent
Hamiltonian of the plain Heisenberg form using the unitary mapping
\begin{eqnarray}\label{utraf}
\tilde S_l^x &&=\cos\varphi_l\,S_l^x-\sin\varphi_l\,S_l^y
\nonumber\\
\tilde S_l^y &&=-\sin\varphi_l\,S_l^x+\cos\varphi_l\,S_l^y
\end{eqnarray}
with $\tilde S_l^z=S_l^z$, $\varphi_l=2l\varphi_0$, and
$\tan(2\varphi_0)=D/J$. Expressed in terms of $\tilde{\bf S}_l$ the
Hamiltonian reads $\tilde H^{(S)}=\sqrt{J^2+D^2}\sum_l
\tilde{\bf S}_l\cdot\tilde{\bf S}_{l+1}$. Now, we note that
higher order terms in $\tilde{\lambda}$ will also contribute to the Raman
operator.  However, following the discussion after (\ref{commu}) it is
obvious that (\ref{utraf}) will not simultaneously reduce $H^{(S)}$ and
$H^{(R)}(\omega_{in,out})$ to a
canonical Heisenberg form.  Therefore, in the new basis, the Raman
operator takes on the form $\tilde H^{(R)}(\omega_{in,out})=\tilde
J_R(\omega_{in,out})\,\tilde H^{(Heis)} +\tilde D_R(\omega_{in,out})
\,\tilde H^{(DM)}+O(\tilde{\lambda}^2)$. The only part of $\tilde
H^{(R)}(\omega_{in,out})$ which does not commute with $\tilde H^{(S)}$
to lowest order in $\tilde{\lambda}$ is
the DM-interaction, i.e.\ $\tilde H^{(R)}(\omega_{in,out})
\equiv \tilde D_R(\omega_{in,out})\,\tilde H^{(DM)}$.

This completes our derivation of the Raman operator for the {\em
homogeneous} phase of NaV$_2$O$_5$ as realized for $T>T_C$. Quite
generally the preceeding demonstrates that a DM-contribution to the Raman
operator $H^{(R)}(\omega_{in,out})$ will occur in multiband systems
whenever a DM exchange-interaction is allowed locally. Obviously it is
tempting to analyse the effects of this form of  $H^{(R)}(\omega_{in,
out})$ also on a {\em dimerized} spin-liquid state, as present in
NaV$_2$O$_5$ for $T<T_C$ and similarly in CuGeO$_3$ for $T<T_{SP}$
\cite{Hase93}. To this end let $H^{(R)}(\omega_{in,out})$ act on a pure
dimer state $|\Phi_0\rangle = |s_1... s_\mu ...\rangle$, where $\mu$
labels nearest-neighbor pairs of spins which are in a relative singlet
state $|s_\mu\rangle$ - for the case of NaV$_2$O$_5$ these pairs of spins
correspond to pairs of rungs $(2l,2l+1)$. One obtains
\begin{eqnarray}\label{con1}
H^{(DM)}|\Phi_0\rangle = 
\textsl{}\sum_\mu (&& -2i|... t_{\mu}^z ...\rangle
 -|... t_{\mu}^xt_{\mu+1}^y ...\rangle 
\nonumber\\
&& + |... t_{\mu}^yt_{\mu+1}^x ...\rangle)~.
\label{HR_dimer}
\end{eqnarray}
Here $|t_{\mu}^\alpha\rangle$ ($\alpha=x,y,z$) refers to triplet states on
the dimer-bonds. While the 2nd and 3rd term on the rhs.\ of (\ref{con1})
comprise of the usual total-spin zero, two-magnon excitation, the first
term refers to a single-triplet state of {\em only} $z$-direction. This
shows that single-magnon Raman-excitations are allowed in the presence of
the DM-interaction. A single-magnon Raman line of this type has a clear
experimental signature: it should show {\it no splitting} in an external
magnetic field parallel to ${\bf e}$ (here along $z$) and it should split
into {\it two} branches for a field perpendicular to the DM-vector.  To
our knowledge this signature has not yet been observed in experiment.


\section{CONCLUSIONS}

Motivated by recent ESR experiments \cite{Luther98,Lohmann00} which
probe the existence of a DM interaction in NaV$_2$O$_5$, we have
presented a microscopic study of the possible impact of this interaction
on the Raman process. We have derived the Raman operator in the
homogeneous state of NaV$_2$O$_5$ and, additionally, have discussed
its effect in the dimerized state.
 
In the dimerized state two Raman-modes have been observed in NaV$_2$O$_5$
in bb-polarization at 66cm$^{-1}$ and 104cm$^{-1}$. Tentatively these
modes have been ascribed to magnetic bound states of total-spin
zero.\cite{Lemmens} On the other hand, for $T<T_C$ INS displays two well
defined magnon-excitations, the energies of which, if properly zone-folded
to zero momentum coincide with the aforementioned two Raman
modes\cite{Gros99,Yosihama98}. Yet, Raman experiments show no indication
of a splitting of these modes in an external magnetic field.  We therefore
conclude that the Raman-modes should result from a two-magnon processes
(see (\ref{HR_dimer})).

Clear evidence for a DM-vector in NaV$_2$O$_5$ along the $z$-direction has
been provided by ESR experiments.\cite{Lohmann00} While these authors have
interpreted there findings in terms of quasi-static charge-fluctuations
above $T_C$, we believe, in view of the results presented here, that such
an interpretation of the ESR-data is not necessary. In fact, the
ESR-experiments can be understood in terms of the local DM-vector present
{\it also} in the high-temperature phase.

In conclusion we have pointed out, that a local DM-vector gives rise to a
non-trivial DM-contribution to the magnetic Raman-process whenever at
least two {\it non-equivalent} exchange paths exist between the two
magnetic moments considered. We have presented an explicit evaluation of
this DM-contribution to the Raman-operator for the case of the
quarter-filled ladder compound NaV$_2$O$_5$ and we have shown, that one-
and two-magnon processes arise naturally within this scenario. We have
obtained estimates for the exchange-coupling constant along the
$b$-direction in good agreement with experiment. Moreover we have
evaluated the spin-orbit coupling constant within our cluster-approach.
Finally, we note that evidence for DM-interactions in the two-dimensional
dimer-compound SrCu$_2$(BO$_3$)$_2$\cite{Kageyama99} have been found by
ESR \cite{Nojiri99} and far-infrared spectroscopy\cite{Room99}. Therefore
one might speculate if one-magnon Raman modes with the special signature
described in the previous section could be observable in
SrCu$_2$(BO$_3$)$_2$.


\section{ACKNOWLEDGMENTS}

We thank P. Lemmens and P. van Loosdrecht for stimulating discussions. We
would like to acknowledge  the support of the German Science Foundation
and the hospitality of the ITP in Santa Barbara where part of this work
was carried out.  This research was supported by the National Science
Foundation under Grant No. PHY94-07194.

\begin{appendix} 
\newcommand{\deu}[1]{d^\dagger_{1#1\uparrow}}
\newcommand{\dzu}[1]{d^\dagger_{2#1\uparrow}}
\newcommand{\ded}[1]{d^\dagger_{1#1\downarrow}}
\newcommand{\dzd}[1]{d^\dagger_{2#1\downarrow}}
\newcommand{\deuv}[1]{d^{\phantom{\dagger}}_{1#1\uparrow}}
\newcommand{\dzuv}[1]{d^{\phantom{\dagger}}_{2#1\uparrow}}
\newcommand{\dedv}[1]{d^{\phantom{\dagger}}_{1#1\downarrow}}
\newcommand{\dzdv}[1]{d^{\phantom{\dagger}}_{2#1\downarrow}}
\newcommand{\pu}[1]{p^\dagger_{#1\uparrow}}
\newcommand{\pd}[1]{p^\dagger_{#1\downarrow}}
\newcommand{\puv}[1]{p^{\phantom{\dagger}}_{#1\uparrow}}
\newcommand{\pdv}[1]{p^{\phantom{\dagger}}_{#1\downarrow}}
\section{Exchange integral and Raman amplitude}
In this appendix we present details of the evaluation of the 2, 3,
and 4-particle eigenstates on a rung as well as the matrix elements
relevant to the exchange integral and the Raman operator. We begin with
the 3-particle space on rung $l$, which, in the subspace of no
double-occupancy of the $d$-levels and total-spin $z$-component
$S_z=\uparrow$ can be created by
\begin{eqnarray}\label{ap1}
\begin{array}{lll}
3'_{1l\uparrow}=\deu{l}\dzd{l}\pu{l} & ~ &
3'_{2l\uparrow}=\ded{l}\dzu{l}\pu{l} \\
3'_{3l\uparrow}=\dzu{l}\pu{l}\pd{l} & ~ & 
3'_{4l\uparrow}=\deu{l}\pu{l}\pd{l}\\
3'_{5l\uparrow}=\deu{l}\dzu{l}\pd{l}
\end{array}
\end{eqnarray}
The set of corresponding states with $S_z=\downarrow$, is obtained by
reversing $\uparrow$ to $\downarrow$ for each operator without
changing their relative order. Diagonalizing the rung-Hamiltonian in
this sector yields a (spin degenerate-)ground state
$|3_{0l\uparrow(\downarrow)}\rangle$ with energy $E_{30}$
\begin{eqnarray}\label{ap2}
|3_{0\uparrow(\downarrow)}\rangle = &&
a \left(
|3'_{1\uparrow(\downarrow)}\rangle+|3'_{2\uparrow(\downarrow)}\rangle]
+ b [|3'_{4\uparrow(\downarrow)}\rangle-|3'_{3\uparrow(\downarrow)}\rangle
\right)
\nonumber\\
&&-2a|3'_{5\uparrow(\downarrow)}\rangle
\nonumber\\
E_{30}=&& \frac{1}{2}(3\epsilon_p - t_\perp - \epsilon) 
\end{eqnarray}
For brevity the site index '$l$' has been suppressed and
\begin{eqnarray}\label{ap3}
a = &&-\sqrt{2t^2_{pd}/[12 t_{pd}^2 + (\epsilon-\epsilon_p+t_\perp)^2]} 
\nonumber \\
b = &&a(\epsilon-\epsilon_p+t_\perp)/(2t_{pd}) 
\nonumber\\
\epsilon= &&\sqrt{12t_{pd}^2 + (\epsilon_p-t_\perp)^2}
\end{eqnarray} 
For $\epsilon_p\approx -3eV$, $t_{pd}\approx -1eV$, and $t_\perp\approx
0.25eV$ one gets $a\approx -0.16$, $b\approx 0.65$, and $E_{30}=-7eV$.

In the 2-particle space with no double-occupancy of the $d$-levels
there are thirteen states, the creation operators of which we label as
follows
\begin{eqnarray}\label{ap4}
\begin{array}{lll}
2'_{1l}   =\deu{l}\dzd{l}&
2'_{2l}   =\ded{l}\dzu{l}&
2'_{3l}   =\ded{l}\dzd{l}\\  
2'_{4l}   =\deu{l}\dzu{l}& 
2'_{5l}   =\ded{l}\pu{l} &
2'_{6l}   =\deu{l}\pd{l} \\
2'_{7l}   =\deu{l}\pu{l} &
2'_{8l}   =\ded{l}\pd{l} & 
2'_{9l}   =\dzu{l}\pd{l} \\
2'_{10l}  =\dzd{l}\pu{l} &
2'_{11l}  =\dzd{l}\pd{l} &
2'_{12l}  =\dzu{l}\pu{l} \\
2'_{13l}  =\pu{l} \pd{l} &
\end{array}
\end{eqnarray}
To simplify matters we will consider the high-energy states with two
holes on the oxygen site as decoupled from the remaining
Hilbert space. In the following these states will be discarded when
evaluating the exchange integrals. With this simplification the
eigenstates $|2_{il}\rangle$ are created by the following set of
operators
\begin{eqnarray}\label{ap5}
2_{1}   =&& 2'_{1} \nonumber\\
2_{2}   =&& 2'_{2} \nonumber\\
2_{3}   =&& 2'_{3} \nonumber\\
2_{4}   =&& 2'_{4} \nonumber\\
2_{5}   =&& \frac{1}{\sqrt{2}}(2'_{7}-2'_{12}) \nonumber\\
2_{6}   =&& \frac{1}{\sqrt{2}}(2'_{8}-2'_{11}) \nonumber\\
2_{7}   =&& \frac{1}{2}(2'_{5}+2'_{6}-2'_{9}-2'_{10}) \nonumber\\
2_{8}   =&& \frac{1}{2}(2'_{5}-2'_{6}+2'_{9}-2'_{10}) \nonumber\\
2_{9}   =&& \frac{1}{\sqrt{2}}(2'_{7}+2'_{12}) \nonumber\\
2_{10}  =&& \frac{1}{\sqrt{2}}(2'_{8}+2'_{11}) \nonumber\\
2_{11}  =&& \frac{1}{2}(2'_{5}+2'_{6}+2'_{9}+2'_{10}) \nonumber\\
2_{12}  =&& (2'_{5}-2'_{6}-2'_{9}+2'_{10}+\beta_2 2'_{13})\beta_3 \nonumber\\
2_{13}  =&& (2'_{6}-2'_{5}-2'_{10}+2'_{9}+\gamma_2 2'_{13})\gamma_3
\end{eqnarray}
where, as before, the site index has been suppressed and
\begin{eqnarray}\label{ap6}
\beta_1(\gamma_1)=&&
\pm\epsilon_p\mp t_\perp+\sqrt{16t_{pd}^2+(\epsilon_p-t_\perp)^2}
\\
\beta_2(\gamma_2)=&&8t_{pd}/\beta_1(\gamma_1)
\nonumber\\
\beta_3(\gamma_3)=&&\beta_1(\gamma_1)/\left[8t_{pd}\sqrt{1+
\frac{\beta_1(\gamma_1)^2}{16t^2_{pd}}} ~\right]
\nonumber
\end{eqnarray}
with upper(lower) signs on the r.h.s of (\ref{ap6}) referring to
$\beta(\gamma)$. The eigenenergies are given by
\begin{eqnarray}\label{ap7}
\begin{array}{llll}
E_{2_{12}} &\equiv E_{20} &= 2\epsilon_p-\beta_1/2 &\approx -6.95eV\\
E_{2_5}=...=E_{2_8} &\equiv E_{21} &= \epsilon_p-t_\perp &\approx -3.25eV\\
E_{2_9}=...= E_{2_{11}} &\equiv E_{22} &= \epsilon_p+t_\perp
&\approx -2.75eV\\
E_{2_{13}}&=E_{23} &\equiv 2\epsilon_p+\gamma_1/2 &\approx -1.80eV\\
E_{2_1}=...=E_{2_4} &\equiv E_{24} &= 0 \\                           
\end{array}
\end{eqnarray}
and have been labeled into ascending order of their numerical values
as relevant to NaV$_2$O$_5$.

The 3-particle space is fourfold degenerate with respect to $H_0$ and
the eigenstates are created by
\begin{eqnarray}\label{ap8}
\begin{array}{ll}
4'_{1l}   =\deu{l}\dzu{l}\pu{l}\pd{l}&
4'_{2l}   =\ded{l}\dzu{l}\pu{l}\pd{l}\\
4'_{3l}   =\deu{l}\dzd{l}\pu{l}\pd{l}&
4'_{4l}   =\ded{l}\dzd{l}\pu{l}\pd{l}
\end{array}
\end{eqnarray}
where $E_{40}=2\epsilon_p$.

To second order in $H_1$, the exchange integral $J$ is obtained from the
energy-dependent transverse spin-flip matrix-elements  $J^{-+}(z)$ and
$J^{+-}(z)$ of the corresponding second-order effective Hamiltonian
\begin{eqnarray}\label{ap10}
\frac{1}{2}J^{-+}(z)
&&\equiv\langle 3_{0l\downarrow}3_{0l+1\uparrow}|W
\frac{1}{z-H_0}
W|3_{0l\uparrow}3_{0l+1\downarrow}\rangle 
\end{eqnarray}
where  $W$ stands for $H_1$ and the energy variable $z$ is zero in the
evaluation of the exchange integral.  The factor $1/2$ in front of
$J^{-+}$ corresponds to the fact that in $H^{(Heis)}$ $S_l^x S_{l+1}^x +
S_l^y S_{l+1}^y = \frac{1}{2}(S_l^+S_{l+1}^- + S_l^-S_{l+1}^+)$.

To second order the Raman scattering amplitude is obtained by considering
Eq.\ (\ref{ap10}) again, however with $W$ denoting the current operator,
i.e. $j_b/e$, in this case and with $z$ depending on the energy of the
incoming/outgoing photon $\omega_{in}$/$\omega_{out}$. Then,
\begin{eqnarray}\label{ap101}
R^{-+}(\omega_{in},\omega_{out})=
J^{-+}(\omega_{in})+J^{-+}(-\omega_{out})
\end{eqnarray}
The first term on the r.h.s of the previous equation describes the process
where first the incoming photon is absorbed in going into the intermediate
state, while the second term  describes the process where the intermediate
state is reached by first emitting the outgoing photon.

Equation (\ref{ap10}) is evaluated using first the transition
amplitudes $\langle\mu|H_1|3_{0l\uparrow(\downarrow)}3_{0l+1\downarrow
(\uparrow)}\rangle$ from the 3$\otimes$3-particle ground states into
the bare intermediate 2$\otimes$4-particle states $|\mu\rangle$ as
constructed from (\ref{ap4}) and (\ref{ap8}) and second by projecting the
latter onto the 2$\otimes$4-particle eigenstates of $H_0$, i.e.
(\ref{ap5}) and (\ref{ap8})
\begin{eqnarray}\label{ap12}
J^{-+}(z)&= & 2 \sum_{ij,\mu>6,\nu>6} \left[
\langle 3_{0l\downarrow}3_{0l+1\uparrow}|W|\mu\rangle
\langle\mu|2_{il}4_{jl+1}\rangle\right.
\nonumber \\
&&
\phantom{=\sum_{ij} [}
\left.\frac{
\langle 2_{il}4_{jl+1}|\nu\rangle
\langle\nu|W|3_{0l\uparrow}3_{0l+1\downarrow}\rangle
}{z-(E_{2_i}+E_{4_j}-2E_{30})}\right]
\end{eqnarray}
The preceding involves 2(4)-particle states on rungs $l$($l+1$) only,
since the hermitian-conjugate exchange path involving 4(2)-particle
states on rungs $l$($l+1$) can be accounted for by the global prefactor
of 2, both, for $W=H_1$ and $W=j_b/e$. Moreover, since up to a factor of
$(-i)i\equiv 1$ the weights in the numerator of (\ref{ap12}) are
identical for $W=H_1$ and $W=j_b/e$ we consider the former only. Table
(\ref{ap9}) lists the bare intermediate states $|\mu\rangle$ and the
corresponding weights.
\begin{equation}\label{ap9}
\begin{array}{c|c|c|c}
\mu & |\mu\rangle &
\langle\mu|H_1|3_{0l\uparrow}3_{0l+1\downarrow}\rangle &
\langle\mu|H_1|3_{0l\downarrow}3_{0l+1\uparrow}\rangle \\
\hline
1 & |2'_{1l}4'_{3l+1}\rangle
& -a^2 t_{pp} & a^2 t_{pp} \\
2 & |2'_{2l}4'_{3l+1}\rangle
& -a^2 t_{pp} & a^2 t_{pp} \\
3 & |2'_{2l}4'_{2l+1}\rangle
& -a^2 t_{pp} & a^2 t_{pp} \\
4 & |2'_{1l}4'_{2l+1}\rangle
& -a^2 t_{pp} & a^2 t_{pp} \\
5 & |2'_{3l}4'_{1l+1}\rangle
& 0 & -4a^2 t_{pp} \\
6 & |2'_{4l}4'_{4l+1}\rangle
&  4a^2 t_{pp} & 0 \\
7 & |2'_{6l}4'_{3l+1}\rangle
& a b t_{pp} & a b t^\ast_\downarrow \\
8 & |2'_{5l}4'_{3l+1}\rangle
& 0 & -ab (2t^\ast_\downarrow+t_{pp}) \\
9 & |2'_{6l}4'_{2l+1}\rangle
& a b (2t^\ast_\uparrow+t_{pp}) & 0 \\
10 & |2'_{5l}4'_{2l+1}\rangle
& -a b t^\ast_\uparrow & -a b t_{pp} \\
11 & |2'_{7l}4'_{4l+1}\rangle
& -a b (t^\ast_\downarrow+2t_{pp}) & 0 \\
12 & |2'_{8l}4'_{1l+1}\rangle
& 0 & a b (t^\ast_\uparrow+2t_{pp}) \\
13 & |2'_{10l}4'_{3l+1}\rangle
& +a b t^\ast_\uparrow & a b t_{pp} \\
14 & |2'_{9l}4'_{3l+1}\rangle
& -a b (2t^\ast_\uparrow + t_{pp}) & 0 \\
15 & |2'_{9l}4'_{2l+1}\rangle
& -a b t_{pp} & -a b t^\ast_\downarrow \\
16 & |2'_{10l}4'_{2l+1}\rangle
& 0 & a b (2t^\ast_\downarrow+t_{pp}) \\
17 & |2'_{11l}4'_{1l+1}\rangle
& 0 & -a b (t^\ast_\uparrow+2t_{pp}) \\
18 & |2'_{12l}4'_{4l+1}\rangle
& a b (t^\ast_\downarrow + 2t_{pp}) & 0 \\
19 & |2'_{13l}4'_{3l+1}\rangle
& b^2 t^\ast_\uparrow & -b^2 t^\ast_\downarrow \\
20 & |2'_{13l}4'_{2l+1}\rangle
& -b^2 t^\ast_\uparrow & b^2 t^\ast_\downarrow
\end{array}
\end{equation}
The constraint in (\ref{ap12}) on the summation over the indices $\mu$ and
$\nu$ reflects the restriction to intermediate states with at most one
$p$-hole. Using (\ref{ap5}), (\ref{ap7}), and (\ref{ap9}) it is a matter
of straightforward algebra to show that
\begin{eqnarray}\label{ap11}
J^{-+}(z)=\sum^4_{i=1}\frac{A^{-+}_i}{z-\Delta E_i}
\end{eqnarray}
where
\begin{eqnarray}\label{ap13}
\Delta E_1 =&& E_{20}+E_{40}-2E_{30}\approx 1.05eV \\
\Delta E_2 =&& E_{21}+E_{40}-2E_{30}\approx 4.75eV \nonumber\\
\Delta E_3 =&& E_{22}+E_{40}-2E_{30}\approx 5.25eV \nonumber\\
\Delta E_4 =&& E_{23}+E_{40}-2E_{30}\approx 6.20eV \nonumber
\end{eqnarray}
and
\begin{eqnarray}\label{ap14}
A^{-+}_1 =&& -\frac{b^2(8bt_{pd}+3a\beta_1)^2t^{\ast\,2}_\uparrow}
{2[16t^2_{pd}+\beta_1(\epsilon_p-t_\perp)]}
\\
\approx &&
-0.0246eV^2 - i 0.281eV \tilde{\lambda} + 0.803 \tilde{\lambda}^2
\nonumber\\
A^{-+}_2 =&& 8a^2b^2t^\ast_\uparrow(t^\ast_\uparrow+t_{pp})
\nonumber\\
\approx &&
-0.00504eV^2 - i 0.0133eV \tilde{\lambda} - 0.0886 \tilde{\lambda}^2
\nonumber\\
A^{-+}_3 =&& a^2b^2t^{\ast\,2}_\uparrow
\nonumber\\
\approx &&
0.000339eV^2 + i 0.00388eV \tilde{\lambda} - 0.0111 \tilde{\lambda}^2
\nonumber\\
A^{-+}_4 =&&\frac{b^2(8bt_{pd}-3a\gamma_1)^2t^{\ast\,2}_\uparrow}
{2[16t^2_{pd}-\gamma_1(\epsilon_p-t_\perp)]} 
\nonumber\\
\approx &&
-0.000181eV^2 - i 0.00206eV \tilde{\lambda} + 0.00589 \tilde{\lambda}^2
\nonumber
\end{eqnarray}

Note that the ferro/antiferromagnetic signs of the amplitudes at
$\tilde{\lambda}=0$ are related to the triplet/singlet character of the
intermediate states. Eg., $A^{-+}_3$ corresponds to a matrix element where
the intermediate states are given by $2_9$, $2_{10}$, $2_{11}$, all of
which are triplets, therefore a ferromagnetic sign of $A^{-+}_3$ arises.
Inserting the numerical values of  $A^{-+}_i$ into (\ref{ap11}) we get
$J^{-+}(z=0)\approx 0.049eV-1.492\tilde{\lambda}^2/eV +i
0.542\tilde{\lambda}$.
\end{appendix}

\end{document}